\documentclass[prl,twocolumn,amssymb,amsmath,aps,superscriptaddress,showpacs]{revtex4}
\usepackage{amsfonts}
\usepackage{graphicx}
\usepackage{epsfig}
\def\bea{\begin{eqnarray}}
\def\eea{\end{eqnarray}}
\def\be{\begin{equation}}
\def\ee{\end{equation}}

\begin{document}
\title{Resonant Pair Tunneling in Double Quantum Dots}
\author{Eran Sela and Ian Affleck \date{\today}}
\affiliation{Department of Physics and Astronomy, University of
British Columbia, Vancouver, B.C., Canada, V6T 1Z1}

\begin{abstract}
We present exact results on the non-equilibrium current fluctuations
for 2 quantum dots in series throughout a crossover from non-Fermi
liquid to Fermi liquid behavior described by the 2 impurity Kondo
model. The result corresponds to resonant tunneling of carriers of
charge $2e$ for a critical inter-impurity coupling. At low energy
scales, the result can be understood from a Fermi liquid approach
that we develop and use to also study non-equilibrium transport in
an alternative double dot realization of the 2 impurity Kondo model
under current experimental study.
\end{abstract}
\pacs{75.20.Hr, 71.10.Hf, 75.75.+a, 73.21.La}

\maketitle

\paragraph*{Introduction.}
Measurements of nonequilibrium shot noise in current fluctuations in
electronic devices~\cite{Schottky18} has became a practical tool to
probe strongly correlated systems with elementary excitations whose
charge, $e^*$, possibly differs from the electron charge $e$, the
prominent examples being the observation of the Cooper-pair charge
$e^*=2 e$ in normal metal-superconductor junctions~\cite{Lefloch03},
or fractional charges in quantum Hall samples~\cite{Picciotto97}.
Remarkably, many-body physics with unusual emergent excitations
arises even when the interactions occur in a single point, e.g., in
an impurity in a metal, or in a nanoscale quantum dot (QD)
connecting few leads. Theoretical studies of various quantum
impurity problems encountered the notion of non-Fermi liquid (NFL)
behavior, with a fully inelastic scattering of an incoming electron
into an out-going scattering state which does not include any single
electron component~\cite{Affleck93}. It is of general interest to
study shot noise is QD systems showing such elusive NFL behavior.

Nontrivial effective charges emerge even for quantum  impurity
problems showing regular Fermi liquid (FL) behavior, for example in
the basic single impurity Kondo model, realized by a single magnetic
(Kondo) QD coupled to leads, studies of shot noise~\cite{Mier02}
lead to a prediction of a universal fractional charge~\cite{Sela06}
$e^*=5 e/3$ in the low temperature regime which was detected
experimentally~\cite{Zarchin08}, reflecting a combination of single
electron and two-electron backscattering. The crossover which is
typically addressed in experiment~\cite{Delattre09} is very rarely
understood theoretically. In this paper we find that a simple and
yet unusual ``noninteracting-like" picture for transport of
particles with effective charge $e^*=2e$ emerges along an entire
crossover from NFL to FL behavior occurring in double QDs in
series~\cite{Georges99,Izumida,Sela09} exhibiting the physics of the
2-impurity Kondo model (2IKM).

The simplest 2IKM consists of two impurity spins $(S_L,S_R)$,
coupled to two channels of conduction electrons and interacting with
each other through an exchange interaction $K$. After the standard
``unfolding transformation"~\cite{Affleck91}, reducing the two
spin-$1/2$ channels to four chiral Dirac fermions, $\psi_{i
\alpha}(x)$, $i=1,2=L,R$, $\alpha=\uparrow,\downarrow$, $x \in
\{-\infty,\infty \}$, the Hamiltonian becomes $H = H_0+H_K$ where
$H_0 =\sum_{j , \alpha} \int dx \psi^\dagger_{j \alpha} i \partial_x
\psi_{j \alpha}$ and
\begin{eqnarray}
\label{eq:HK} H_K= J_L (\psi_L^\dagger \vec{\sigma} \psi_L) \cdot
\vec{S}_L +J_R (\psi_R^\dagger \vec{\sigma} \psi_R)\cdot \vec{S}_R
+K \vec{S}_L \cdot \vec{S}_R,
\end{eqnarray}
where $\vec{\sigma}$($\vec{\tau}$) is a vector of Pauli matrices
acting in spin(channel) space. For this model a NFL quantum critical
point (QCP) was found at $K = K_c \sim T_K$~\cite{Jones88}
separating a local singlet FL phase at $K>K_c$ from a Kondo screened
FL phase at $K < K_c$. However, more realistic models containing
inter-channel tunneling,
\begin{equation}
\label{eq:HPS} H_{PS} = V_{LR}  \psi^\dagger_L \psi_R + {\rm{H.c.}}
,
\end{equation}
[or, $H_{PS}= {\rm{Re}}V_{LR} (\psi^\dagger \tau^1
\psi)-{\rm{Im}}V_{LR} (\psi^\dagger \tau^2 \psi)$ with implicit sum
over spin and channel indices] do not show a critical
point~\cite{Sakai,Fye94}. The reason for this is that
Eq.~(\ref{eq:HPS}) results in a relevant perturbation with dimension
$1/2$ at the QCP~\cite{Affleck92}, leading to an energy scale $T^*
\propto T_K |\nu V_{LR}|^2 + (K-K_c)^2/T_K$ which is finite even at
$K=K_c$, below which an effective FL theory takes over. Here $\nu$
is the density of states of the conduction electrons. This crossover
from NFL to FL behavior is reflected in the conductance of double
QDs~\cite{Georges99,Zarand06,Izumida,Sela09}, and in particular
geometries, e.g., the series geometry, we were able to calculate it
exactly~\cite{Sela09}. However this information is not sufficient to
uncover the nature of the transport.

In this paper we study the full counting
statistics~\cite{Levitov93,Gustavsson06} (FCS) for charge transfer
through a series double QD along the full NFL to FL crossover. In
general, charge is transferred in units of $e$ or $2e$. A peculiar
situation occurs at $K \to K_c$, where $2e$ becomes the basic charge
unit along the full crossover. This striking behavior is not
captured in a slave-boson mean field calculation~\cite{Lopez04}.

We also derive a local Fermi liquid Hamiltonian governing the
physics  below  $T^*$. Using an enhanced understanding of this
crossover we go beyond previous
works~\cite{Yamada79,Schlottmann84,Jones89}
 in determining all coupling
constants in this effective Hamiltonian and obtain a universal
theory depending only on an energy scale, $T^*$, similar to
Nozi\`{e}res FL theory (FLT) for the single impurity
problem~\cite{Nozieres74}, and on the new FL boundary condition
associated with the ratio $|K-K_c| /(\nu  |V_{LR}| T_K)$. This
approach helps to understand the charge $2e$ carriers.

In the geometry proposed by Zarand \emph{et al.}~\cite{Zarand06},
where transport proceeds between two leads connected via one QD side
coupled to a second QD coupled to another lead, exact results on the
crossover are not available. Nonetheless, we use our FLT to
calculate universal non-equilibrium transport and noise properties
at low energies when the NFL critical behavior is destabilized by a
non-zero $K-K_c$. Our predictions can be probed
experimentally~\cite{GG08}.

\paragraph*{Full counting statistics.} We will obtain the full charge transfer distribution
in a series double QD tuned to the 2IKM regime, along the crossover
from NFL to FL behavior, using the formulation we developed in
Ref.~\cite{Sela09}.

In terms of abelian bosonization one can write the original free
fermion theory with $H_K \to 0$ and $H_{PS} \to 0$ in terms of 8
chiral Majorana fermions $\chi_i ^A$, $\chi_1
^A=\frac{\psi_A^\dagger+\psi_A}{\sqrt{2}}$, $\chi_2
^A=\frac{\psi_A^\dagger-\psi_A}{\sqrt{2}i}$, associated with the
real $(i=1)$ and imaginary $(i=2)$ parts of the charge, spin, flavor
and spin-flavor fermions $(A = c,s,f,X)$; for a definition of these
fermions, see Ref.~\cite{Sela09}. Then the free Hamiltonian is
$H_0[\{\chi'\}] = \frac{i}{2} \sum_{j=1}^8\int dx \chi'_j
\partial_x \chi'_j$, where $\{\chi'\}=\{  \chi_2^X,\chi_1^f,\chi_2^f,\chi_1^X,\chi_1^c,\chi_2^c,\chi_1^s,\chi_2^s
\}$. The Fermi operator $\psi_f$ gives rise to charge $e$ (and no
spin) tunneling from left to right, and changes $Y = (N_L-N_R)/2$ by
$1$, $N_i$ being the total fermion number in lead $i=L,R$.

Turning on $H_K$, the QCP is obtained at $K=K_c$ from the free case
by a change in boundary condition (BC) occurring only for the first
Majorana fermion, $\chi_1(0^-) = - \chi_1(0^+)$. For energies $\ll
T_K$, the leading terms in the Hamiltonian describing deviations
$K-K_c$ as well as finite $V_{LR}$ can be written~\cite{Sela09} in a
new basis $\{\chi\}$, where $\chi_1(x) = \chi_1'(x)~ {\rm{sgn}}(x)$
and $\chi_i = \chi'_i$, $(i=2,\ldots,8)$, as $H_{QCP}= H_0[\{\chi\}]
+ \delta H_{QCP}$ where~\cite{remark1}
\begin{eqnarray}
\label{eq:HQCP} \delta H_{QCP}=i\sum_{i=1}^2  \lambda_i\chi_i(0) a.
\end{eqnarray}
Here $a$ is a local Majorana fermion, $a^2=1/2$, and
\begin{equation}
\label{eq:lambda12} \lambda_1 =c_1
\frac{K-K_c}{\sqrt{T_K}},~~~\lambda_2 =c_2 \sqrt{T_K}  |\nu V_{LR}|,
\end{equation}
where $c_1$ and $c_2$ are constant factors of order 1. Those
couplings determine two energy scales $\lambda_1^2$, $\lambda_2^2$,
and the total crossover scale is $\lambda^2=\lambda_1^2 +
\lambda_2^2 = T^*$. The operators in $\delta H_{QCP}$ have scaling
dimension $1/2$, hence they destabilize the QCP; below the crossover
scale $T^* =\lambda_1^2 + \lambda_2^2$ the system flows to FL fixed
points whose nature depend on the ratio $\lambda_1/\lambda_2$.

By definition the FCS is obtained from the cumulant generating
function $\chi(\mu)$ for the probability distribution function
$P(Q)$ to transfer $Q$ units of charge during the waiting time
$\mathcal{T}$ (which is sent to infinity), $\chi(\mu)=\sum_Q e^{i Q
\mu}P(Q)$. The cumulants $\langle \delta^n Q \rangle$ can be found
from $ \langle \delta^n Q \rangle = (-i )^n
\frac{\partial^n}{\partial \mu^n} \ln \chi (\mu) \big|_{\mu=0}$. In
fact, due to a formal equivalence of our non-equilibrium formulation
and that of Schiller and Hershfield~\cite{Schiller98} for a single
QD tuned to the Toulouse limit, we can borrow directly the results
of Gogolin and Komnik for the FCS for that model~\cite{Gogolin06};
translating between the parameters of the two models in the limit
$T^*  \ll T_K$, we obtain
\begin{eqnarray}
\label{eq:chi} \frac{\ln \chi(\mu)}{\mathcal{T}}= \int_{-
\infty}^{\infty} \frac{d \epsilon}{4 \pi} \ln  \bigl[ 1
 +\sum_{n=-2}^{2} A_n(\epsilon) \cdot (e^{ i \mu n}-1)\bigr].
\end{eqnarray}
Here $A_1(\epsilon)=\frac{2 \lambda^2_1 \lambda^2_2 }{4 \epsilon^2+
\lambda^4} [n_F(1-n_L)+n_R(1-n_F)] $,
$A_2(\epsilon)=\frac{\lambda^4_2 }{4 \epsilon^2+ \lambda^4} n_L
(1-n_R)$, $A_{-n} = A_n|_{L \leftrightarrow R}$, $n_F =
(1+e^{\epsilon/T})^{-1}$, $n_{L,R} = n_F(\epsilon \mp eV)$. The
presence of one particle as well as two particle transport processes
in our model is apparent from the $\mu$ dependence of the two terms
$\propto (e^{\pm i \mu }-1)$ and $\propto (e^{\pm 2 i \mu }-1)$ in
Eq.~(\ref{eq:chi}), respectively. At $K = K_c$, giving $A_1 =
A_{-1}=0$, Eq.~(\ref{eq:chi}) is equivalent to the formula for the
FCS of spinless noninteracting fermions of charge $2e$ transmitted
though a resonant level of width $\sim T^*$, namely the
noninteracting formula~\cite{Levitov93,Gogolin06} is obtained from
Eq.~(\ref{eq:chi}) by the replacement $2 \mu \to \mu$, $n_{L,R} \to
n_F(\epsilon \mp eV/2)$ and adding an overall factor of 2.

The emergence of two particle resonant tunneling at $K = K_c$
follows from Eq.~(\ref{eq:HQCP}). In this case $\delta H_{QCP}
=\frac{i}{\sqrt{2}} \lambda_2 (\psi^\dagger_f+\psi_f) a$ has the
form of a Majorana resonant level~\cite{Schiller98}. This operator
changes $Y$ by $\pm 1$, while a noninteracting resonant level $d$
with $\delta H  = \lambda_2 (\psi_L^\dagger+\psi_R^\dagger)d +
{\rm{H.c.}}$ changes $Y$[$=(N_L-N_R)/2$] by $\pm 1/2$. In both
models, transport is given by processes of even order in
$\lambda_2$, giving $\Delta Y$ integer in the resonant level model
but $\Delta Y$ even-integer in the 2IKM.  In the limit of large $V$,
in either non-interacting resonant level model or 2IKM, one can do
perturbation theory in $\lambda_2$. The $T=0$ conductance, in this
limit, is given by~\cite{Schiller98,Sela09} $G \propto
\lambda_2^4/V^2$, implying a second order process. This follows
since the first order tunneling between a fermionic state [either
$\psi_f$ or $\psi_i$ $(i=L,R)$ in the two models] with energy of
order $eV$, and the zero energy level ($a$ or $d$) does not conserve
energy. In the small $V$ limit of the 2IKM we can understand the
charge 2e using the FLT developed below.

\begin{figure}[h]
\begin{center}
\includegraphics*[width=70mm]{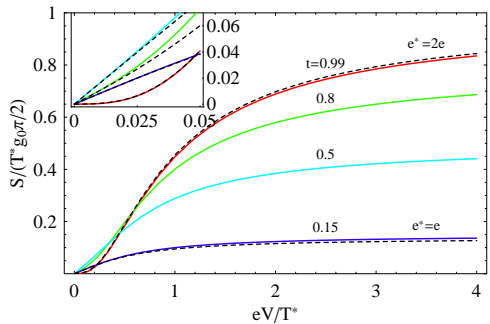}
\caption{Shot noise versus voltage for several values of $K-K_c$
(determined by $t$) at $T=0$. Data is fitted in dashed lines using
Eq.~(\ref{eq:fit}). The inset blows up the FL region, with fits of
$e^* = e$ for ($t = 0.15, 0.5,0.8$) and $e^* =2 e$ for $t=0.99$ ($K
\to K_c$, red curve). \label{fg:q}}
\end{center}
\end{figure}

The two-particle processes can be probed by looking simultaneously
at the current $I=e \langle \delta^1 Q \rangle /\mathcal{T}$ and
noise $S=2e^2 \langle \delta^2 Q \rangle /\mathcal{T}$.
Eq.~(\ref{eq:chi}) gives the $T=V=0$ conductance $G =\frac{dI}{dV}=
g_0 t$ where $t = \lambda_2^2 / \lambda^2 = \frac{|\nu
V_{LR}|^2}{|\nu V_{LR}|^2+(c_1/c_2)^2 (\frac{K-K_c}{T_K})^2} $ and
$g_0 =2 e^2/h$ (or setting $\hbar = 1$ as in the rest of the paper,
$g_0 =e^2/\pi$). Using Eq.~(\ref{eq:chi}), in Fig.~\ref{fg:q} we
plot shot noise $S(V)$ along the crossover from FL ($eV \ll T^*$) to
NFL ($eV \gg T^*$) regimes for various values of $K-K_c$ (determined
by $t$). Experiments~\cite{Zarchin08} extract effective charges by
fitting shot noise measurements with the formula
\begin{equation}
\label{eq:fit} S_{{\rm{fit}}} = 2 e^* g_0 \int _0^{V} t(V')[1-t(V')]
dV',
\end{equation}
where $t(V)$ is extracted from nonlinear conductance measurements
$t(V) = \frac{1}{g_0} \frac{dI}{dV}$, $t(0)  =t$, and $e^*$ is the
effective charge. For $K \ne K_c$ ($t <
 1$), $S_{{\rm{fit}}}$ with
$e^* = e$ gives a good fit for sufficiently small $V$; see inset of
Fig.~\ref{fg:q}. When $|K-K_c| \gg T_K |\nu V_{LR}|$ ($t \ll 1$),
this fit with $e^* = e$ becomes reasonably good along the full
crossover; see curve with $t=0.15$ in Fig.~\ref{fg:q}. For $K \to
K_c$ ($t \to 1$) the fit with $e^* = e$ works only for an extremely
small range $eV \ll \sqrt{T^*/T_K} |K-K_c| \to 0$, and, remarkably,
the \emph{full} curve fits with $S_{{\rm{fit}}}$ with $e^* = 2e$;
see curve with $t = 0.99$.

\paragraph*{Fermi liquid theory.}
While the above results were obtained formally from a calculation in
terms of refermionization, it is desirable to understand them from
an effective interacting theory written in terms of the original
fermions. The reader may skip the technical details and go directly
to the FL Hamiltonian, Eq.~(\ref{eq:FLJJ}).

The crucial observation is that only the linear combination
$(\lambda_1 \chi_1+ \lambda_2 \chi_2)/\lambda$ of the 8 Majorana
fermions at the QCP participates in this crossover described by
Eq.~(\ref{eq:HQCP}). It can be shown that the effect of $\delta
H_{QCP}$ is to modify the BC for this linear combination by a simple
sign change. In order to write down the FL fixed point Hamiltonian,
we define a new basis with modified BC, $\{\eta\}$, where $\eta_1 =
(\lambda_1 \chi_1+ \lambda_2 \chi_2)/\lambda~ {\rm{sgn}}(x)$,
$\eta_2 = (-\lambda_2 \chi_1+ \lambda_1 \chi_2)/\lambda$, and
$\eta_i = \chi_i$, $(i=3,\ldots,8)$. Hence we can write the
Hamiltonian for the FL fixed points as $H_{FL} =
H_0[\{\eta\}]+\delta H_{FL}$. We expect the local interaction
$\delta H_{FL}$ to involve uniquely $\eta_1$, which is the only
field participating in the crossover in Eq.~(\ref{eq:HQCP}). The
only candidate for the leading FL operator with scaling dimension 2
is
\begin{equation}
\label{eq:FL8} \delta H_{FL} =\lambda_{FL} i   \eta_1
\partial_x \eta_1 |_{x=0},
\end{equation}
with $\lambda_{FL} \propto 1/{T^*}$. Comparison of the scattering
phase shift for Eqs.~(\ref{eq:HQCP}) and (\ref{eq:FL8}) actually
gives $\lambda_{FL} = 4/ T^*$.

Equation~(\ref{eq:FL8}) for the FL interaction may be written in
terms of fields which are simply related to the original fermions.
The latter satisfy a FL BC parameterized by a unitary S-matrix,
\begin{equation}
\label{eq:s} \psi_{i \alpha}(0^-) = s_{i j}[\alpha] \psi_{j
\alpha}(0^+) ~~~(\rm{repeated~indices~summed}).
\end{equation}
We will express $H_{FL}$ in terms of single particle scattering
states $\Psi_{i \alpha}$ incoming from channel (lead) $i =1,2=
(L,R)$ with spin $\alpha = \pm 1$,
\begin{equation}
\label{eq:scattering}
 \Psi_{j \alpha}(x) = \theta(x) \psi_{j \alpha}(x)+\theta(-x) s_{j j'}[\alpha] \psi_{j'
 \alpha}(x),
\end{equation}
satisfying $\Psi_{i \alpha}(0^+) = \Psi_{i \alpha}(0^-)$. In our
left moving convention, the region $x>0$ ($x<0$) corresponds to the
incoming(outgoing) part of the field.

To find the S-matrix we should relate the BC of the $\eta$'s,
$\eta_i(0^+) = \eta_i(0^-)$, $(i=1,8)$, to the BC of the $\psi_{i
\alpha}$'s, Eq.~(\ref{eq:s}). The representation of the $\psi_{i
\alpha}$'s in terms of the $\eta$'s (or the $\chi'$'s or $\chi$'s)
is fairly complicated, however quadratic forms of those different
fermions are linearly related. In particular, consider $J_M =:
\psi^\dagger M \psi :$ where $M$ acts in the channel space. It is
straightforward to find the coefficients $c^M_{ij}$ such that $J_M=
\sum_{i,j=1}^8 c_{ij}^M :\chi'_i ~\chi'_j:$. Now consider $J_{M}(x)$
at $x = 0^+$. Using these linear relations together with the BC for
the $\chi'$'s (which depends on $\lambda_1/ \lambda_2$), one can
find $M'$ as function of $M$ and $\lambda_1/ \lambda_2$, such that
$J_{M'}(0^-) = J_{M}(0^+)$. This relation between $M$ and $M'$ can
be used to find the S-matrix, since Eq.~(\ref{eq:s}) implies $M' =
s^\dagger M s$. Using this scheme, starting with $H_{PS}$ with real
$V_{LR}$ we obtain: $s[\alpha] = \cos (2 \delta) - i \alpha \sin (2
\delta) \tau^1$, where $\cos 4 \delta =
\frac{\lambda_1^2-\lambda_2^2}{\lambda_1^2+ \lambda_2^2}$. The last
equation gives $2 \delta$ modulo $\pi$. However, under the
transformation $V_{LR} \to -  V_{LR}$ in Eq.~(\ref{eq:HPS}), we have
$2 \delta \to 2\pi - 2 \delta$~\cite{remark1}. Thus
\begin{equation}
\label{eq:delta}  2\delta = \arg (\lambda_1+i\lambda_2).
\end{equation}
$\delta$, which is denoted as the phase shift, changes from $0$ to
$\pi/2$ as function of $K$, and it takes the value of $\pi/4$ at
$K=K_c$. This first exact result for the phase shift agrees with the
numerical results of Jones~\cite{Jones88}. To obtain the S-matrix
for complex $V_{LR}$, one can start with a phase rotation of the
$\psi$'s, rendering $V_{LR}$ real~\cite{remark1}, and in the end,
transform back to the original basis leads to a rotation of the
S-matrix $s \to \exp (i \tau^3 \arg(V_{LR})/2) s \exp (- i \tau^3
\arg(V_{LR})/2)$.

The FL interaction Eq.~(\ref{eq:FL8}) can be explicitly written in
terms of the the scattering states $\Psi$. To achieve this formally
we (i) switch from quadratic derivative forms of the $\chi's$,
$\mathcal{O}_{kl} =\chi_k i \partial_x \chi_{l}+ \chi_l i \partial_x
\chi_{k}$, in Eq.~(\ref{eq:FL8}), to quartic forms, using $i
\chi_{j}
\partial_x \chi_{j}+ i \chi_{j'} \partial_x \chi_{j'}=:\chi_j \chi_{j'}::\chi_{j'}
\chi_{j}:$, $(j \ne j' )$; (ii) express $:\chi_i \chi_j:$ linearly
in terms of quadratic forms of the $\Psi's$ (these linear relations
have the same coefficients $c^M_{ij}$). The result is $H^{FL}_0
=\int dx {\Psi}^\dagger i
\partial_x \Psi$ (indices~summed), and~\cite{remarkOii}
\begin{eqnarray}
\label{eq:FLJJ} \delta H_{FL}&=& \frac{\cos^2(2
\delta)\mathcal{O}_{11}+\sin^2(2 \delta) \mathcal{O}_{22}+\sin(4
\delta)\mathcal{O}_{12}}{2 T^*} \big|_{x=0},\nonumber \\
\mathcal{O}_{11} &=&\frac{16}{3}(\vec{J}_L^2 + \vec{J}_R^2)
-4(\vec{J}_L + \vec{J}_R)^2 ,\nonumber
\\
 \mathcal{O}_{22}&=&
(J_L-J_R)^2- : \Psi^\dagger_{L \alpha} \epsilon_{\alpha \beta}
\Psi^\dagger_{L \beta} \Psi_{R \gamma} \epsilon_{\gamma \delta}
\Psi_{R \delta}+ {\rm{H.c.}}:,\nonumber
\\
\mathcal{O}_{12} &=&i \sum_{j=L,R} :\Psi^\dagger_{L \alpha}
\epsilon_{\alpha \beta} \Psi^\dagger_{R \beta} \Psi_{j \gamma}
\epsilon_{\gamma \delta} \Psi_{j \delta}:+{\rm{H.c.}},
\end{eqnarray}
where $J_i = :\Psi^\dagger_i \Psi_i:$, $\vec{J}_i = \Psi_i^\dagger
\frac{\vec{\sigma}}{2} \Psi_i$ ($i=L,R$) and $\epsilon_{\alpha
\beta}$ is the antisymmetric tensor. This is the main result of this
section. It gives an explicit form of the interactions between the
scattering states, related to the original electrons by
Eq.~(\ref{eq:scattering}). This interaction is weak in the FL regime
allowing to apply perturbation theory in $1/T^*$. This universal FL
Hamiltonian follows by strong restrictions due to a large symmetry
emerging close to the QCP~\cite{Affleck92} and leading to the simple
form of $H_{QCP}$ in Eq.~(\ref{eq:HQCP}). In practice the symmetry
at the QCP is reduced by marginal and irrelevant operators such as
the leading irrelevant operator $
\partial_x \chi_1 a$ (at the QCP); however they will be associated with
a small parameter $1/T_K$, and hence are neglected at low energies
for $T^* \ll T_K$. For finite $V_{LR}$, $K-K_c$, or intra-lead
potential scattering $V_L \psi^\dagger_L \psi_L+V_R \psi^\dagger_R
\psi_R$, additional marginal and irrelevant terms are produced at
the QCP, part of which were present before. However close enough to
the QCP and starting with a weak coupling problem, namely for $|\nu
V_{LR}|,|\nu V_{i}| \ll 1$ ($i=L,R$), and $|K-K_c| \ll T_K$, those
perturbations can be safely ignored.

The emergence of the basic transport charge $2e$ for $K=K_c$
($\delta = \pi/4$) in the series geometry follows at low energies as
the only term in Eq.~(\ref{eq:FLJJ}) which does not conserve the
number of $\Psi_L$ fermions minus the number of $\Psi_R$ fermions is
$\propto ( \Psi^\dagger_{L \alpha} \epsilon_{\alpha \beta}
\Psi^\dagger_{L \beta})( \Psi_{R \gamma} \epsilon_{\gamma \delta}
\Psi_{R \delta})+ {\rm{H.c.}}$. It converts a spin singlet pair of
$\Psi_L$ fermions into a spin singlet pair of $\Psi_R$ fermions (and
vise versa). At $K=K_c$ the scattering states $\Psi_L$($\Psi_R$) are
waves propagating freely to the right(left), hence the basic
transport mechanism is an inelastic $2e$ \emph{backscattering}.

To calculate transport properties for the Zarand \emph{et al.}
double QD geometry~\cite{Zarand06} which is under current
experimental study~\cite{GG08}, it is necessary to calculate the
single particle Green's function in the 2IKM. Because the electron
field cannot be expressed in terms of the Majorana fermions, we have
not been able to calculate this throughout the crossover, results
being necessarily restricted to the vicinity of the NFL or FL
critical points. Here we consider this system at $K$ slightly
different than $K_c$ in the FL regimes, $T,eV \ll T^*$, where our
FLT can be applied, ignoring particle-hole breaking ($V_{LR}=0$).
The conductance of this system can be expanded as
$G_{{\rm{singlet}}}=g_0[\sin^2 \delta_1 + \beta_1 \{ (T/T^*)^2
+\alpha (eV/T^*)^2\} ]$, and $G_{{\rm{screened}}}=g_0[\cos^2
\delta_1 - \gamma_1 \{ (T/T^*)^2 +\alpha' (eV/T^*)^2\} ]$, where
$\delta_1$ is a small phase shift associated with marginal potential
scattering operators. We assume parity symmetry of the device. Using
Eq.~(\ref{eq:FLJJ}), after a lengthy but straightforward
calculation, we determine universal relations: $\beta_1 = \gamma_1 =
O(1)$, $\alpha = \alpha' = 9/10 \pi^2$. We also calculate the shot
noise in the FL regime, and define effective charges $(e^*/e) = S /
2 I$ for the local singlet FL regime ($K>K_c$), and $(e^*/e)' = S /
2 (g_0 V - I)$ in the Kondo screened phase ($K<K_c$), defined in the
limit $\delta_1 \to 0$. Using Eq.~(\ref{eq:FLJJ}) we obtain $(e^*/e)
= (e^*/e)' = 11/9$. Our predictions should be contrasted with the
measurements on single QDs with $\alpha =3 / 2
\pi^2$~\cite{Grobis08} and $e^*/e =5/3 $~\cite{Zarchin08}.

We thank J. Malecki and Y. Oreg for very helpful discussions. This
work was supported by NSERC (ES $\&$ IA) and CIfAR (IA).

\end{document}